\documentclass[journal]{IEEEtran}

\usepackage{amsmath,amsfonts,amssymb}
\usepackage{array}
\usepackage{url}
\usepackage{graphicx}
\usepackage{subcaption} 
\usepackage{float}
\usepackage{cite}
\usepackage{hyperref}
\usepackage{adjustbox}
\usepackage{bm}  

\usepackage{booktabs}
\usepackage{multirow}
\usepackage[dvipsnames]{xcolor}  

\begin{document}

\title{Towards Ultra-Low-Power Neuromorphic Speech Enhancement with Spiking-FullSubNet}

\author{
  Xiang~Hao\textsuperscript{*},
  Chenxiang~Ma\textsuperscript{*},
  Qu~Yang,
  Jibin~Wu,~\IEEEmembership{Member,~IEEE, }%
  and Kay~Chen~Tan,~\IEEEmembership{Fellow,~IEEE}%
\thanks{\textsuperscript{*}Xiang Hao and Chenxiang Ma contributed equally to this work. Corresponding Author: Jibin Wu (jibin.wu@polyu.edu.hk).}%
  \thanks{Xiang Hao and Chenxiang Ma are with the Department of Computing, The Hong Kong Polytechnic University, Hong Kong SAR, China.}%
  \thanks{Kay Chen Tan is with the Department of Data Science and Artificial Intelligence, The Hong Kong Polytechnic University, Hong Kong SAR, China.}%
  \thanks{Jibin Wu is with the Department of Data Science and Artificial Intelligence, and Department of Computing, The Hong Kong Polytechnic University, Hong Kong SAR, China.}%
  \thanks{Qu Yang is with the Department of Electrical and Computer Engineering, National University of Singapore, Singapore.}
}



\maketitle

\begin{abstract}
Speech enhancement is critical for improving speech intelligibility and quality in various audio devices. In recent years, deep learning-based methods have significantly improved speech enhancement performance, but they often come with a high computational cost, which is prohibitive for a large number of edge devices, such as headsets and hearing aids. This work proposes an ultra-low-power speech enhancement system based on the brain-inspired spiking neural network (SNN) called Spiking-FullSubNet. Spiking-FullSubNet follows a full-band and sub-band fusioned approach to effectively capture both global and local spectral information. To enhance the efficiency of computationally expensive sub-band modeling, we introduce a frequency partitioning method inspired by the sensitivity profile of the human peripheral auditory system. Furthermore, we introduce a novel spiking neuron model that can dynamically control the input information integration and forgetting, enhancing the multi-scale temporal processing capability of SNN, which is critical for speech denoising. Experiments conducted on the recent Intel Neuromorphic Deep Noise Suppression (N-DNS) Challenge dataset show that the Spiking-FullSubNet surpasses state-of-the-art methods by large margins in terms of both speech quality and energy efficiency metrics. Notably, our system won the championship of the Intel N-DNS Challenge (Algorithmic Track), opening up a myriad of opportunities for ultra-low-power speech enhancement at the edge. Our source code and model checkpoints are publicly available at \href{https://github.com/haoxiangsnr/spiking-fullsubnet}{github.com/haoxiangsnr/spiking-fullsubnet}.

\end{abstract}


\section{Introduction}
\IEEEPARstart{M}{icrophones} inevitably pick up various interferences from the surrounding environments, such as ambient noise, which can drastically degrade the quality of perceived speech signals. This prevalent issue calls for speech enhancement (SE) techniques in real-world applications like headsets, hands-free communication, teleconferencing, and hearing aids~\cite{van2009speech, Gannot2017ACP}, etc. Furthermore, SE can also benefit a variety of downstream tasks, such as automatic speech recognition~\cite{kinoshita_reverb_2013}, speaker recognition/verification~\cite{taherian2020robust}, and speech diarization~\cite{ahmad2020speech}. It serves as a crucial front-end processing step to improve the robustness of these systems against signal degradation.

Early SE methods, such as spectral subtraction~\cite{loizou2007speech} and Wiener filtering~\cite{Gannot2017ACP}, were developed to improve speech signals. However, these methods often face challenges in real-world conditions, particularly when dealing with low signal-to-noise ratio (SNR) scenarios~\cite{xu2013experimental,xu2014regression}.
During the last decade, deep learning techniques, such as Long Short-Term Memory (LSTM)~\cite{Fedorov2020TinyLSTMsEN,hao_fullsubnet_2021,hao_fast_2023}, Convolutional Neural Network (CNN)~\cite{hu_dccrn_2020,wang_tf-gridnet_2023}, and Transformer~\cite{xiong22_interspeech}, have improved the SE performance by leaps and bounds~\cite{xu2013experimental,xu2014regression}.
While these deep learning-based SE models offer superior performance, their high computational cost and latency can be prohibitive for deployment on ubiquitous resource-constrained edge devices, such as headsets and hearing aids~\cite{Fedorov2020TinyLSTMsEN, Choi2021RealTimeDA}.

Recently, brain-inspired Spiking Neural Networks (SNNs) have received increasing attention as energy-efficient alternatives to the prevailing deep learning models~\cite{roy2019towards,chen2023hybridcoding,progressivetandem2021,yan2023CQ,zhu2023Ultra-High,yao2023attentionsnn,hu2023FastSNN}.
SNNs employ spike trains to encode and convey information, closely mimicking the operations of biological neural networks~\cite{roy2019towards}. This spike-based communication leads to asynchronous, event-driven computation, where information processing is triggered solely by the arrival of incoming spikes. Furthermore, spiking neurons exhibit rich neuronal dynamics that are believed to be crucial for information processing in the brain~\cite{birkhoff1927dynamical}. This spike-based communication, coupled with the inherent temporal dynamics of spiking neurons, enables efficient and effective temporal signal processing using SNNs. 
Notably, when deployed on emerging neuromorphic chips, SNN models can yield orders of magnitude improvements in energy efficiency and reduced latency compared to conventional artificial neural networks (ANNs) used in deep learning ~\cite{akopyan2015truenorth,davies2018loihi,pei2019towards}.

The superior energy efficiency, low latency, and ability to effectively process temporal signals position SNNs as a compelling approach for performing SE on resource-constrained devices.
This potential has been notably recognized in the recent Intel Neuromorphic Deep Noise Suppression (N-DNS) Challenge~\cite{timcheck_intel_2023}. However, the development of high-performance SNNs for the SE task currently faces several key challenges. One primary challenge arises from the inherent complexity of speech signals, which exhibit temporal variations across multiple time scales. Current SNN models, commonly equipped with simplified spiking neurons like the Leaky Integrate-and-Fire (LIF) model, often face difficulties in effectively handling the high temporal complexity present in speech signals~\cite{gerstner2002spiking}. Moreover, achieving state-of-the-art (SOTA) performance on par with or surpassing conventional ANN systems while ensuring real-time audio streaming, as required in practical applications, demands a holistic system design. Such a design necessitates not only enhanced spiking neuron models, but also the development of effective enhancement workflows and training techniques tailored to the unique characteristics of SNNs. To the best of our knowledge, no existing study has effectively addressed these challenges. 

In this work, we introduce a real-time neuromorphic SE system that
demonstrates competitive denoising performance while exhibiting significantly improved energy efficiency compared to ANN methods. 
First of all, drawing inspiration from recent advancements in deep-learning-based SE research works~\cite{hao_fullsubnet_2021,wang_tf-gridnet_2023,chen22c_interspeech,xiong22_interspeech}, we propose a novel SNN-based SE model named \textit{Spiking-FullSubNet}. This model effectively integrates both full-band and sub-band information, allowing it to capture both global and local spectral characteristics. Specifically, the full-band component receives inputs covering the entire frequency partition, enabling it to capture the global spectral structure of the speech signal. On the other hand, each sub-band component focuses exclusively on a specific frequency band, facilitating effective modeling of local spectral structures. Moreover, to enhance the computation efficiency of the sub-band components, we introduce a brain-inspired frequency band partitioning method. Specifically, motivated by the frequency sensitivity profile of the peripheral auditory system, we employ a finer granularity for low-frequency bands and a coarser granularity for high-frequency bands. 

Furthermore, we propose a novel spiking neuron model called Gated Spiking Neuron (GSN) to enhance the temporal signal processing capability of spiking neurons. Unlike existing spiking neuron models where input information decays exponentially over time, the GSN model dynamically controls the integration of input information and the forgetting of historical information. This unique feature enables the GSN to identify and retain crucial temporal information that is essential for speech enhancement. In summary, our main contributions are threefold:

\begin{itemize}
\item We propose \textit{Spiking-FullSubNet}, a novel real-time neuromorphic SE model that combines recent advancements in speech enhancement and neuromorphic computing. This cross-disciplinary approach significantly enhances both speech enhancement performance and energy efficiency.
\item We propose a novel spiking neuron model called GSN. Unlike existing spiking neuron models that passively filter input information, the GSN model dynamically controls the input information integration and forgetting, thereby facilitating effective temporal information processing that is critical for speech enhancement. 
\item We conducted comprehensive experiments on the Intel N-DNS Challenge dataset to evaluate the proposed neuromorphic SE model. Our model not only demonstrates exceptional speech enhancement capabilities, but also showcases a remarkable improvement in energy efficiency, surpassing SOTA ANN models by almost three orders of magnitude.
\end{itemize}


\begin{figure}
  \centering
  \includegraphics[width=\linewidth]{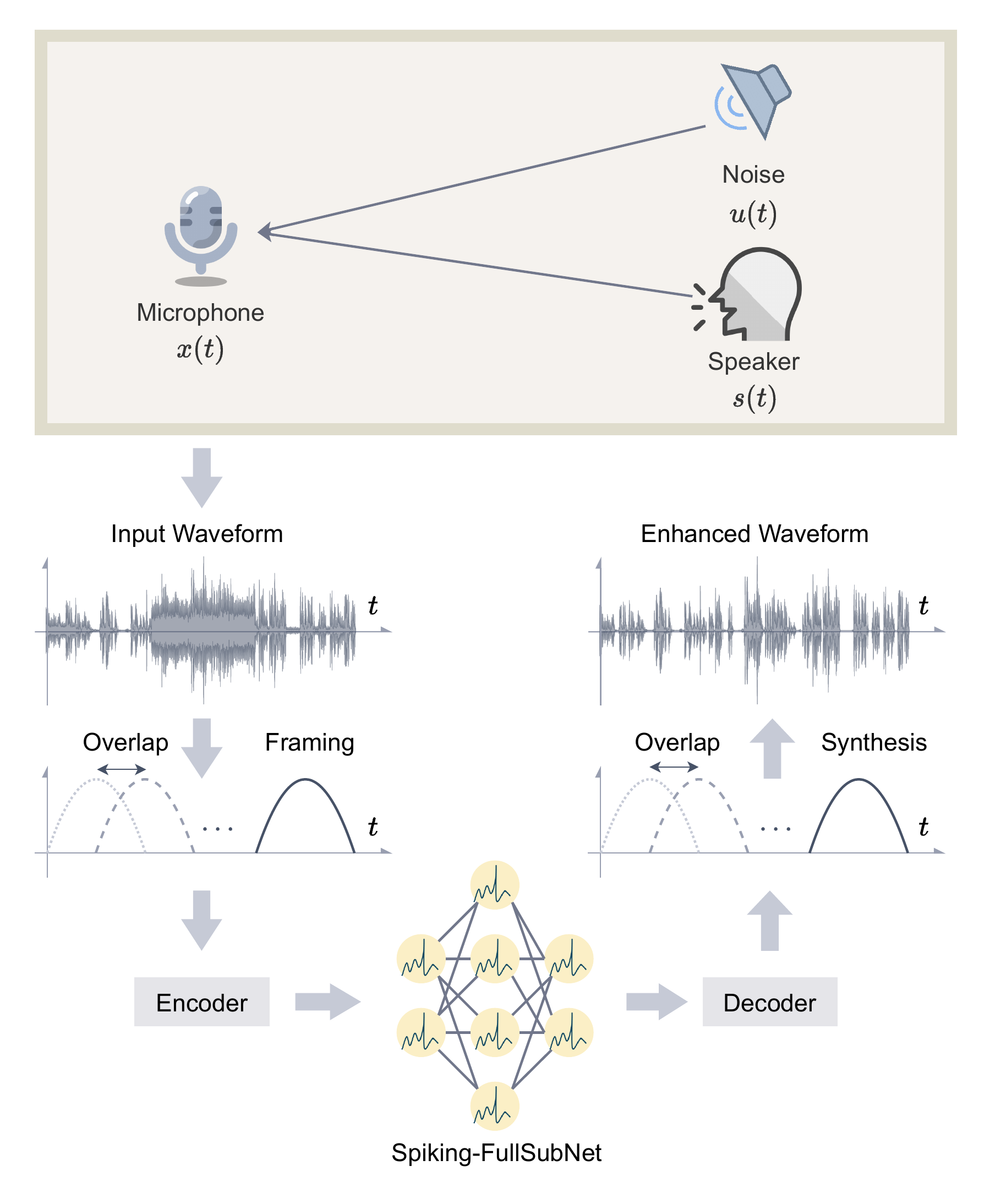}
  \caption{Block diagram of the proposed real-time neuromorphic speech enhancement system.}
  \label{fig:real-time-processing}
\end{figure}

\section{Related Works}
\label{sec:related_works}

\subsection{Spiking Neuron Model}
Among various spiking neuron models introduced in neuroscience~\cite{izhikevich2003simple,hodgkin1952quantitative}, the Leaky Integrate-and-Fire (LIF) model~\cite{gerstner2002spiking} is the most commonly employed one for constructing large-scale neuromorphic computing systems.
Despite their simplicity, LIF neurons have limitations in handling temporal signals with complex structures due to their oversimplified neuronal dynamics. To overcome this limitation, several new spiking neuron models have been introduced recently. For example, the parametric LIF (PLIF) model~\cite{fang2021incorporating} incorporates learnable time constants, allowing historical information to decay at different rates. The adaptive LIF (ALIF) model~\cite{yin2021accurate} adjusts the firing threshold after each spike, effectively achieving firing rate adaptation. This endows spiking neurons with context-dependent processing capability. In a similar vein of research, the GLIF model~\cite{yao2022glif} selectively regulates input currents integration, membrane decaying factors, and neuronal state resetting to enrich neuronal dynamics. These regulations are achieved through trainable parameters along the temporal dimension. While the GLIF model exhibits flexibility in temporal processing, it encounters difficulties when dealing with signals of varying time lengths. This issue is particularly relevant to the speech enhancement task addressed in this paper, which involves handling signals with variable durations.

In addition to incorporating adaptive state variables into spiking neurons to enhance their temporal processing capability, recent studies have introduced multi-compartment neuron models. These models aim to enrich neuronal dynamics by considering the complex morphology of biological neurons. For instance, the TC-LIF model~\cite{zhang2023tc} incorporates two compartments representing soma and dendrites, enabling the simulation of interactive dynamics between these distinct neuronal structures. Building upon this concept, PMSN~ model\cite{chen2024a} incorporates multiple compartments for multi-scale temporal information processing. However, despite these advancements, existing spiking neuron models still face challenges in effectively capturing and retaining essential temporal information, limiting their ability to process signals with complex temporal structures. 

\subsection{Speech Enhancement}
Speech enhancement aims to improve speech intelligibility and quality~\cite{loizou2007speech, Gannot2017ACP}, which need to remove noise from noisy signals captured by microphones as shown in Fig.~\ref{fig:real-time-processing}. Existing methods can be divided into time-domain and frequency-domain approaches.
Time-domain methods~\cite{pandey2019new,pandey2019tcnn} directly estimate the clean speech signal, bypassing spectral analysis and waveform synthesis, while frequency-domain methods~\cite{xu2013experimental,xu2014regression,cao_cmgan_2024,hu_dccrn_2020,hao_fullsubnet_2021,hao_fast_2023} estimate the spectrogram of the clean speech and then convert it back to the time-domain signal. Between these two approaches, frequency-domain methods have received significant research attention, primarily due to the sparse nature of speech in the frequency domain.
Specifically, they can be broadly categorized into spectral magnitude-only enhancement and complex spectrum enhancement.
Spectral magnitude-only methods~\cite{xu2013experimental,xu2014regression} focus on estimating the magnitude of the clean speech spectrum, utilizing the noisy phase for reconstructing the time-domain signal. On the other hand, complex spectrum enhancement methods~\cite{cao_cmgan_2024,hu_dccrn_2020,hao_fullsubnet_2021,hao_fast_2023} estimate both the real and imaginary parts of the complex spectrum, which have exhibited greater potential in enhancing speech quality by leveraging the full spectral information.

\subsection{Sub-band Modeling in Speech Enhancement}
In recent years, there has been a notable shift in research focus towards the utilization of sub-band modeling in both single-channel and multi-channel SE~\cite{hao_fullsubnet_2021,hao_fast_2023,wang_tf-gridnet_2023,chen22c_interspeech,xiong22_interspeech}.
In contrast to traditional full-band modeling, sub-band modeling involves the separation of input audio into multiple frequency bands, which are then processed independently. Specifically, each sub-band model takes in a noisy sub-band signal along with its adjacent frequency bands, and then predicts the corresponding clean sub-band signal. 
This method leverages the distinct stationary characteristics of speech and noise. Speech signals are inherently non-stationary, exhibiting dynamic and variable properties over time. Conversely, many types of noise are relatively stationary, meaning their statistical properties remain more consistent and stable~\cite{gerkmann2011unbiased,li2016non}. 
In addition, sub-band modeling focuses on the local spectral pattern presented in the current and neighboring frequencies, which has been proven informative for discriminating between speech and other signals~\cite{diethorn2004subband,lin2003subband}.
Furthermore, the sub-band models are also effective in modeling the reverberation as the room's reverberation time (RT60) is frequency-dependent~\cite{kuttruff2016room}. 

However, the sub-band modeling approach comes with a trade-off. While it leverages the distinct stationary characteristics of speech and noise, it can also result in the loss of the global spectral structure of the speech signal. This global spectral information is also crucial for effective SE. To address this issue, recent works have proposed a full-band and sub-band fusion modeling approach~\cite{hao_fullsubnet_2021,hao_fast_2023,wang_tf-gridnet_2023,chen22c_interspeech,xiong22_interspeech}. In these works,  a full-band model and several sub-band models are combined, allowing them to complement each other and capture both the local and global spectral information~\cite{wang_tf-gridnet_2023}.
Though these fusion-based methods have led to significant improvements, sub-band modeling can still be computationally costly, as it requires processing each frequency band separately. This poses challenges for real-time edge applications. In this work, we propose a novel approach that applies different granularity levels to various frequency sub-bands, significantly reducing the computational cost while maintaining the speech enhancement performance.

\subsection{Neuromorphic Speech Processing}
SNNs have recently emerged as a promising approach for power-efficient speech processing. Compared to traditional ANNs, SNNs can offer significant advantages in terms of reduced computational complexity.
An early work~\cite{tavanaei2017spiking} employs Izhikevich neurons~\cite{izhikevich2003simple} as feature extractors, which are trained using unsupervised spiking-timing-dependent plasticity (STDP)~\cite{caporale2008spike} to recognize spoken digits. Subsequent works~\cite{wu2018biologically,wu2018spiking} introduce a SOM-SNN framework that incorporates a self-organizing map (SOM) for feature representation, followed by an SNN for pattern classification.
Recent studies leverage deeper SNNs and more advanced learning rules to enhance classification performance. For instance, deep convolutional SNNs coupled with the tandem learning rule have achieved significantly improved performance on keyword spotting tasks~\cite{tandem2021,yang2022deep}. Recurrent networks of spiking neurons~(RSNNs) are also exploited for speech recognition~\cite{zhang2015digital,bellec2018long}, holding enhanced memory capacity and bringing improvements over the feedforward counterparts. These earlier studies focus on small vocabulary speech recognition tasks. More recently, Wu et al.~\cite{wu2020deep} apply deep SNNs to large vocabulary continuous speech recognition tasks and demonstrate competitive accuracy compared to ANN-based systems. Notably, a recent benchmark study~\cite{blouw2019benchmarking} compared the performance of neuromorphic keyword spotting systems to ANN-based systems that deployed on conventional computing hardware. The study evaluated metrics such as inference speed, energy cost per inference, and dynamic energy consumption. The findings from this benchmark study suggest that neuromorphic systems can significantly reduce the energy costs per inference while maintaining equivalent inference accuracy compared to their traditional ANN counterparts. However, there is a lack of study of neuromorphic technologies for the speech denoising task.

\begin{figure*}
    \centering
  \begin{subfigure}[b]{0.29\textwidth}
    \centering
    \includegraphics[width=\textwidth]{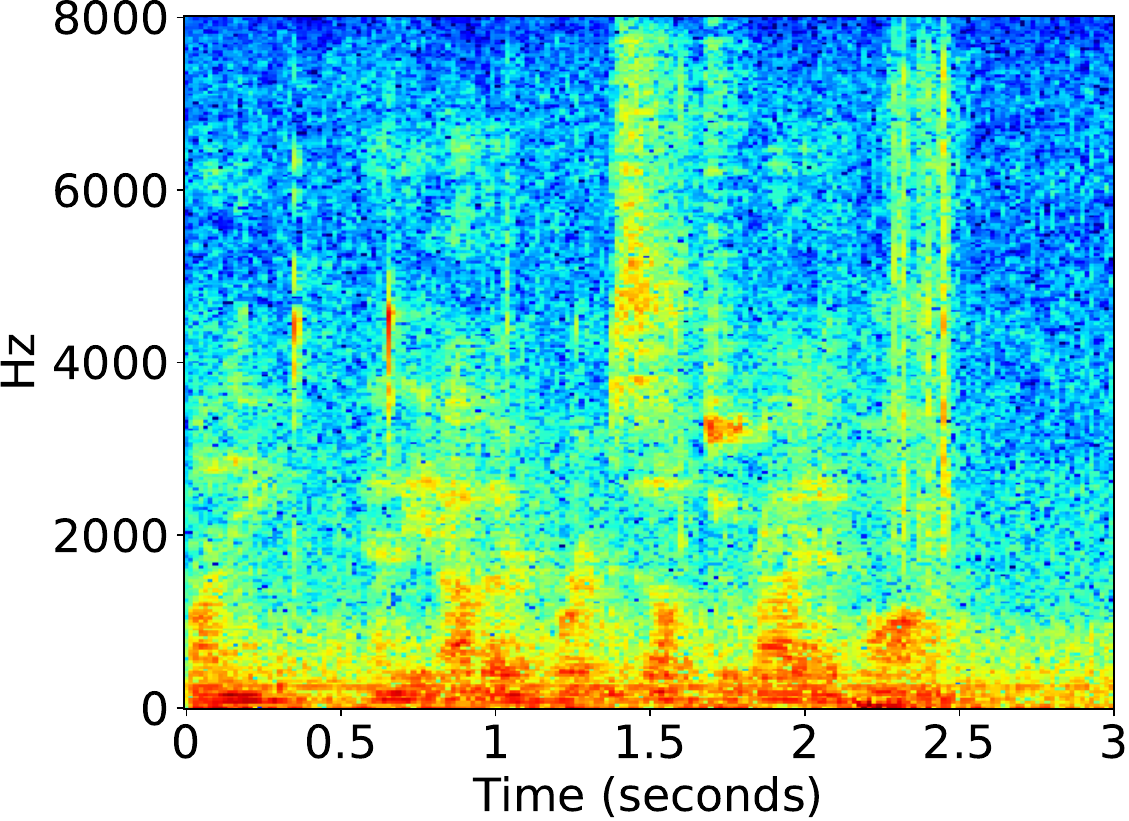}
    \caption{Spectrogram of Noisy Speech Signal}
    \label{fig:gull}
  \end{subfigure}%
  \hspace{0.05\textwidth}
  \begin{subfigure}[b]{0.29\textwidth}
    \centering
    \includegraphics[width=\linewidth]{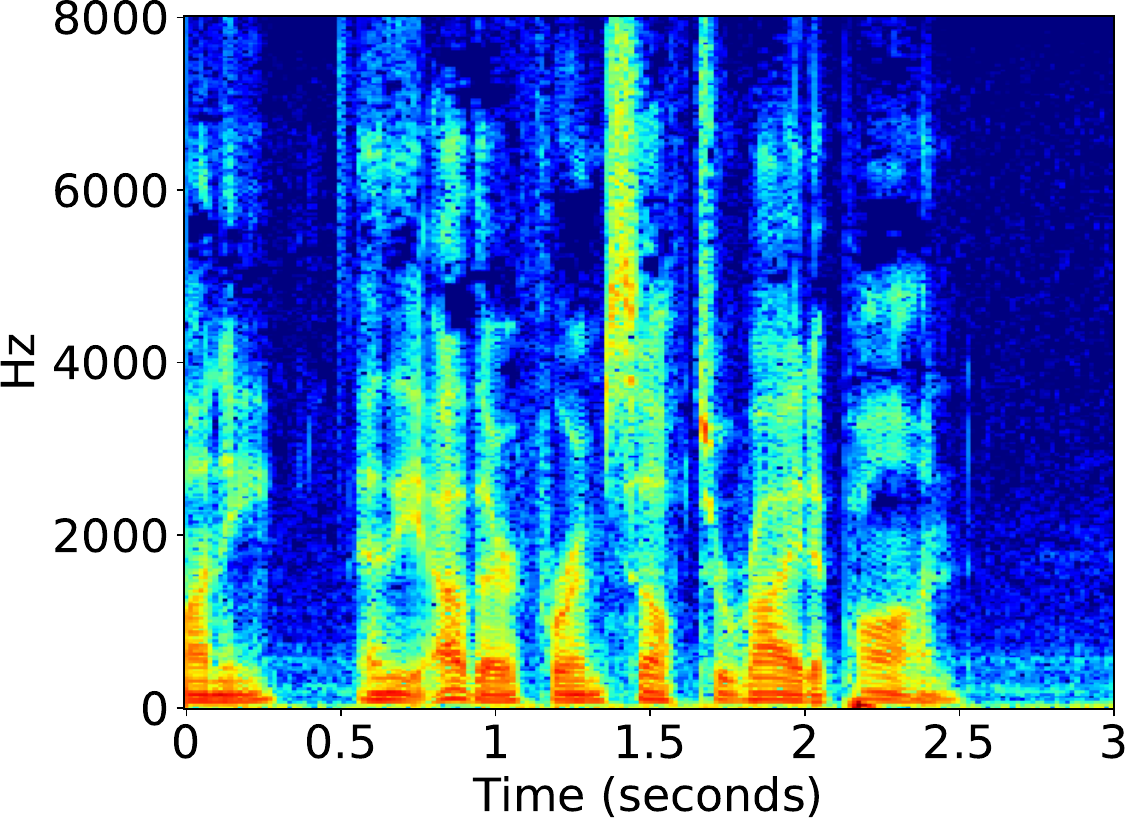}
    \caption{Spectrogram of Clean Speech Signal}
    \label{fig:tiger}
  \end{subfigure}%
  \hspace{0.05\textwidth}%
  \begin{subfigure}[b]{0.29\textwidth}
    \centering
    \includegraphics[width=\linewidth]{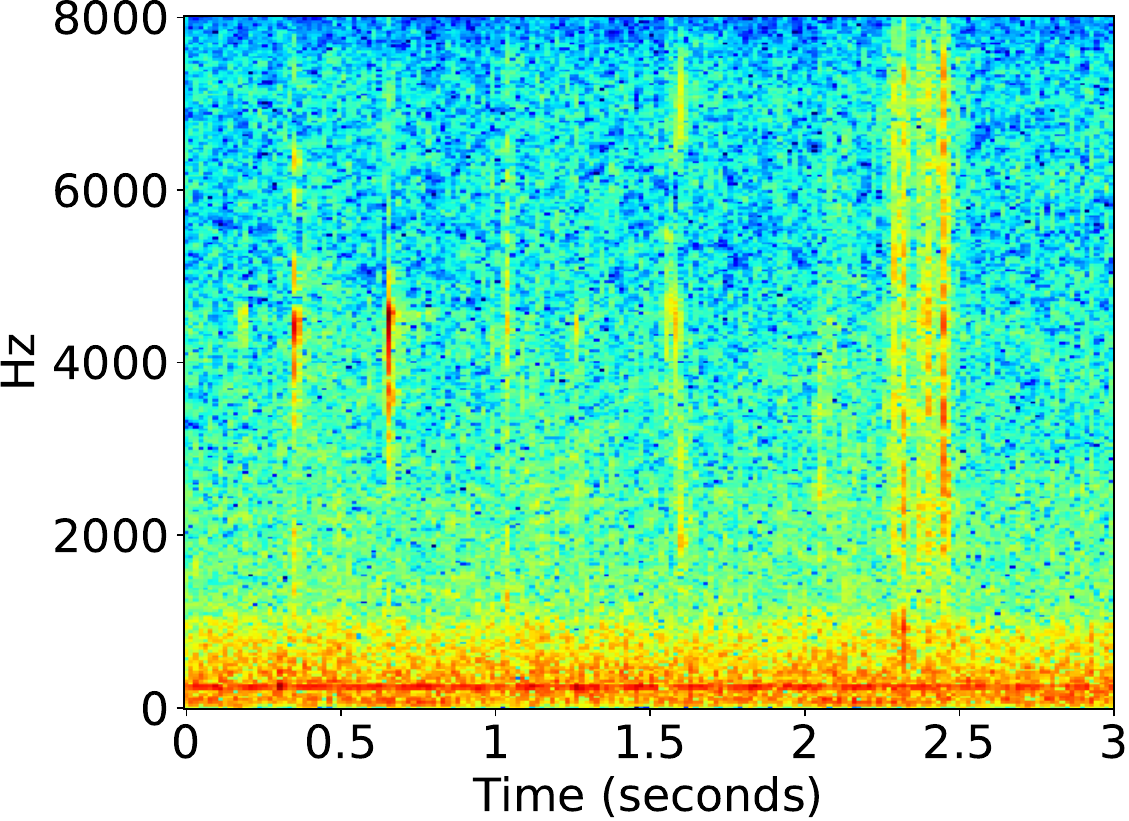}
    \caption{Spectrogram of Noise Signal}
    \label{fig:mouse}
  \end{subfigure}
  \caption{Time-frequency magnitude spectrogram of different signals. The speech enhancement methods aim at recovering a clean speech from a noisy observation by removing the unwanted noise.}
  \label{fig:visual_spec}
\end{figure*}

\section{Background}
\label{sec:background}

\subsection{Spiking Neuron Model}
The most commonly used spiking neuron model is the leaky integrated-and-fire (LIF) neuron~\cite{gerstner2002spiking}. This model is favorable in terms of computational complexity and analytical tractability.
The LIF neuron maintains an internal state known as the membrane potential, which decays over time at a rate determined by the time constant $\tau$. Meanwhile, the neuron integrates the input current. When the membrane potential surpasses the predefined threshold~$\vartheta$, an output spike is generated and transmitted to downstream neurons. This is then followed by a resetting process, where the membrane potential is reset to a specific value. Such neuronal dynamics can be described by the following discrete-time formulation:
\begin{equation}
  \bm{i}^l(t) = \bm{W}^{l}_{mn}\bm{o}^{l-1}(t) + \bm{W}^{l}_{nn}\bm{o}^l(t-1) + \bm{b}^{l}
\end{equation}
\begin{equation}
  \bm{u}^l(t) = \lambda \bm{u}^l(t-1) + \bm{i}^l(t)
\end{equation}
\begin{equation}
  o_i^l(t) = \Theta(u_i^l(t) - \vartheta)=\begin{cases}1 & u_i^l(t) >= \vartheta\\0 & \text{otherwise}\end{cases}
\end{equation}
\begin{equation}
  \bm{u}^l(t) = \bm{u}^l(t)  - \vartheta \bm{o}^l(t)
\end{equation}
where $\bm{W}^{l}_{mn}$ and $\bm{W}^{l}_{nn}$ denote the feed-forward and recurrent weight matrices at the $l^{th}$ layer, respectively,  $\lambda=\exp(-1/\tau)$ controls the decay rate of the membrane potential $\bm{u}^l(t)$, $\bm{i}^l(t)$ represents the input current, and $\bm{b}^l$ is the bias term.

\subsection{Formulation of Speech Enhancement}
In a typical acoustic environment, as depicted in Fig.~\ref{fig:real-time-processing}, the input to a speech enhancement system is a time-varying waveform that undergoes continuous sampling and quantization. The signal captured by the microphone can be represented as
\begin{equation}
  x(t) = s(t) + u(t) \quad {t=1,2,\ldots,T},
\end{equation}
where $s(t)$ refers to the clean speech signal and $u(t)$ denotes a mixture of noise sources.
Most mainstream methods~\cite{hu_dccrn_2020,fu2019metricGAN,zhang_deepmmse_2020,hao_fullsubnet_2021,cao_cmgan_2024} avoid directly enhancing the time-domain signal due to its high non-stationarity and the subsequent analytical challenges. Consequently, the STFT of the microphone signal can be expressed as
\begin{equation}
  x(n,f) = s(n,f) + u(n,f),
\end{equation}
where $n=1,2,\ldots,N$ and $f=1,2,\ldots,F$ represent the time-frame and the frequency-bin indexes, respectively.
For a speech enhancement model, the input feature to the model is $x(n,f)$, with its time-frequency (TF) magnitude spectrogram depicted in Fig.~\ref{fig:visual_spec}(a). The goal is to eliminate unwanted noise, which can be defined as $\hat{s}(n,f)$, with its magnitude spectrogram shown in Fig.~\ref{fig:visual_spec}(b). The noise reference is $u(n,f)$, as shown in Fig.~\ref{fig:visual_spec}(c).

\section{Method}
\label{sec:method}
\begin{figure*}
  \centering
  \includegraphics[width=\textwidth]{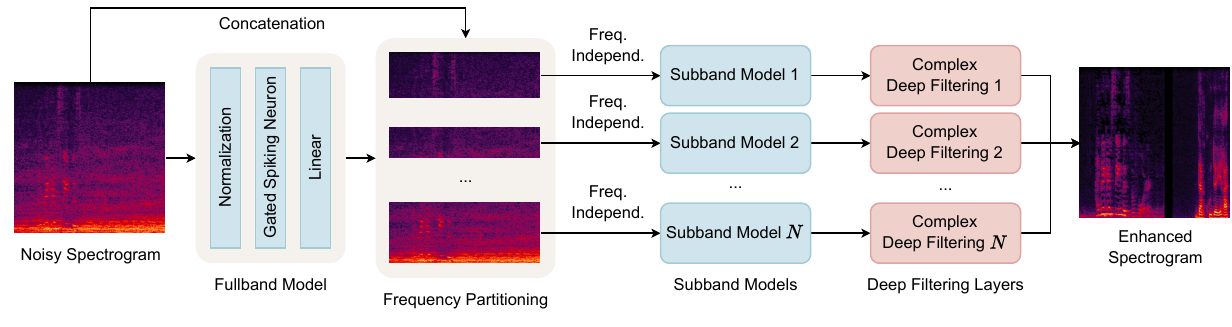}
  \caption{Diagram of the proposed Spiking-FullSubNet architecture. The architecture integrates a full-band model and sub-band models, with gated spiking neurons serving as the core of each model, to effectively enhance noisy speech signals. The full-band model operates on the noisy magnitude spectrogram to capture global spectral patterns, while the sub-band components focus on specific frequency bands to effectively model local spectral information. By incorporating newly proposed GSNs into both the full-band and sub-band models, the temporal processing capability is greatly improved. Finally, deep filtering is employed as the training target to obtain the enhanced spectrogram.}
  \label{fig:workflow}
\end{figure*}

In this section, we elaborate on the proposed Spiking-FullSubNet model, as depicted in Fig.~\ref{fig:workflow}. Firstly, we introduce a novel spiking neuron model called GSN, specifically designed for speech processing. Next, we present the approach of full-band and sub-band fusion for speech denoising that we have adopted in this work. Lastly, we introduce the model training details. 

\subsection{Gated Spiking Neuron}
\label{sec:gsn}
For speech enhancement, the objective is to remove unwanted noises from audio recordings. These noises can originate from various sources, such as background sounds, microphone interference, or transmission distortions, and their signal intensity varies over time. Due to its constant membrane decay rate, the commonly-used LIF model~\cite{gerstner2002spiking} cannot handle these dynamic changes. In these scenarios, the LIF neuron encounters two challenges. It might struggle to filter out noise during periods of high noise intensity effectively, or it may excessively filter informative audio signals during periods of low noise intensity.
These challenges occur as the fixed decay factor either cannot provide sufficient attenuation of the noisy signal, resulting in inadequate enhancement or decays the neuron's potential too quickly, leading to a loss of desired audio contents.

\begin{figure}[!t]
  \centering\includegraphics[width=0.43\textwidth]{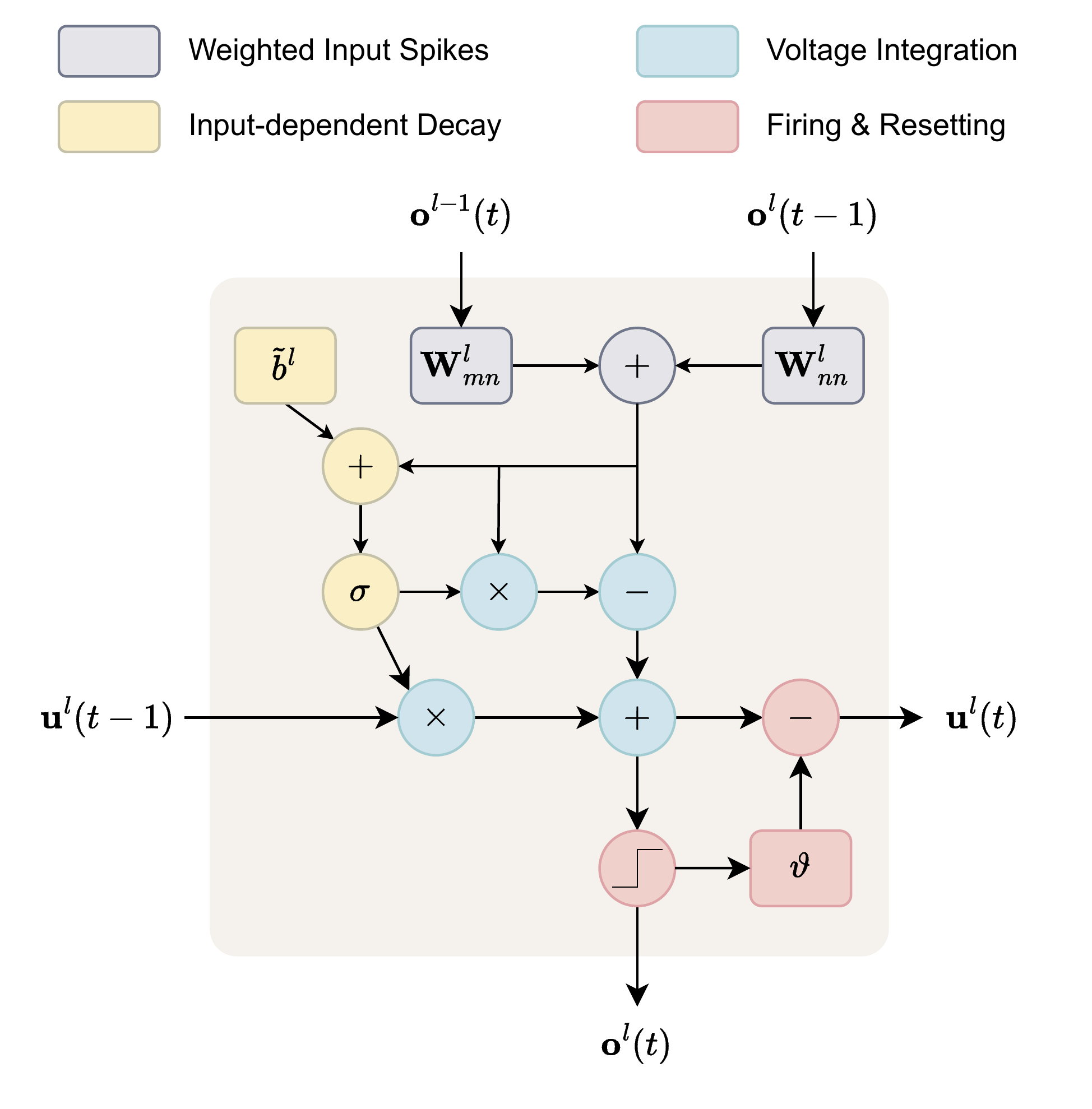}
  \caption{Illustration of the proposed GSN model, which regulates the membrane decay rate at each time step based on the feedforward and recurrent inputs.}
  \label{Fig:GSN}
\end{figure}

A straightforward solution is to adopt different decay factors at each time step, allowing flexible adaptation to the changing noise levels in the input audio signals. However, this would introduce a huge number of parameters, especially for long time duration. Moreover, it would not work well for audio signals with a variable time duration, which are commonly encountered in SE tasks~\cite{Choi2021RealTimeDA}. To address this issue, we propose a novel spiking neuron model called GSN that regulates the decay rate at each time step in an input-dependent manner. The input-dependent decay in our GSN is implemented by modeling the decay as a function of both feed-forward and recurrent input spikes, as shown in Fig.~\ref{Fig:GSN}. A sigmoid function $\sigma(\cdot)$ is applied to constrain the decay rate within the range of $0$ to $1$. The corresponding membrane potential dynamics are formally expressed as:
\begin{equation}
  \bm{i}^l(t) = \bm{W}^l_{mn}\bm{o}^{l-1}(t) + \bm{W}^l_{nn}\bm{o}^l(t-1) + \bm{b}^l,
  \label{eq_current}
\end{equation}
\begin{equation}
  \bm{\lambda}^l(t) = \sigma(\bm{W}^l_{mn}\bm{o}^{l-1}(t) + \bm{W}^l_{nn}\bm{o}^l(t-1) + \bm{\widetilde{b}}^l),
  \label{eq_decayeq}
\end{equation}
\begin{equation}
  \bm{u}^l(t) = \bm{\lambda}^l(t)\bm{u}^l(t-1) + (1 - \bm{\lambda}^l(t))\bm{i}^l(t).
\end{equation}

In order to save model parameters and reduce overall computation, we employ weight-sharing by using the same weight matrices, denoted as $\bm{W}^l_{mn}$ and $\bm{W}^l_{nn}$, in Eqs. (\ref{eq_current}) and (\ref{eq_decayeq}). Notably, the GSN model can adaptively modulate the neuron's membrane potential along the temporal dimension while avoiding the need for a large number of parameters associated with the total time steps. Additionally, GSN bears a resemblance to the forget gate widely used as a critical component of the LSTM architecture~\cite{hochreiter1997long}. However, this temporal gating mechanism has been underexplored in existing spiking neuron models.

\subsection{Spiking-FullSubNet Architecture}

Based on the GSN model introduced in Section~\ref{sec:gsn}, we further develop the Spiking-FullSubNet that can achieve high-performance real-time speech enhancement. Spiking-FullSubNet exploits full-band and sub-band modeling to capture both global (long-distance cross-band dependencies) and local (signal stationarity differences) spectral patterns, respectively. Notably, we propose to employ varying processing granularity for different frequency partitions, mimicking human auditory perception, to enhance the model's processing efficiency. 

\subsubsection{Full-Band Processing}
The full-band model operates on magnitude spectral features extracted from the noisy speech signal. The input feature vector $\mathbf{x}(n)$ on audio frame $n$ is given by
\begin{equation}
  \mathbf{x}(n) = \left[ |x(n, 1)|, |x(n, 2)|, \dots, |x(n, F)| \right]^\top \in \mathbb{R}^{F},
\end{equation}
where $F$ denotes the total number of frequency bins,  $x(n, f)$ represents the complex Fourier coefficient of the frame $n$ and frequency bin $f$, and $|\cdot|$ denotes extracting the magnitude of the complex Fourier coefficient. The sequence of feature vectors across all frames is denoted by $\mathbf{X}$:
\begin{equation}
  \mathbf{X} = \left[ \mathbf{x}(1), \mathbf{x}(2), \dots, \mathbf{x}(T) \right] \in \mathbb{R}^{T \times F},
\end{equation}
where $T$ is the total number of discrete time frames. We stack GSNs to process $\mathbf{X}$ by capturing both the global spectral content and the interactions between frequency bins, yielding a spectral embedding $\mathbf{E}$ of the same dimensions as $\mathbf{X}$, i.e.,  $\mathbf{E} \in \mathbb{R}^{T \times F}$.

\begin{figure}
  \centering
  \includegraphics[width=1\linewidth]{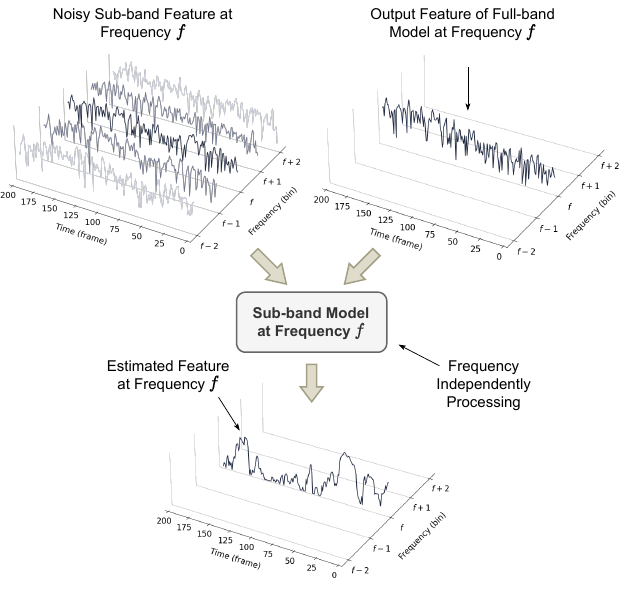}
  \caption{Illustration of the sub-band processing in Spiking-FullSubNet. The input
    is a vector $\mathbf{x}_f(n)$ comprising the magnitude bin of frequency $f$, its $4$ neighboring frequency bins, and the corresponding bin in spectral embedding output from the full-band model.}
  \label{fig:sub_band_processing}
\end{figure}

\subsubsection{Existing Sub-Band Processing Approach}
The sub-band models process each frequency band independently, focusing on the stationarity differences between the speech and noise, local spectral patterns, and reverberation characteristics~\cite{hao_fullsubnet_2021,wang_tf-gridnet_2023}. As shown in Fig.~\ref{fig:sub_band_processing}, for a given time $n$ at frequency $f$, the input to the sub-band model is a vector $\mathbf{x}_f(n)$ comprising the noisy magnitude bin at frequency $f$, its $2 \times N$ neighboring frequency bins, and the corresponding bin in spectral embedding output by the full-band model:
\begin{equation}
  \begin{split}
    \mathbf{x}_f(n) = & [\underbrace{|x(n, f-N)|, \ldots, |x(n, f-1)|}_\text{Lower neighboring frequencies}, \\
        & |x(n, f)|, \mathbf{E}(n, f), \\
        & \underbrace{|x(n, f+1)|, \dots, |x(n, f+N)|}_\text{Higher neighboring frequencies} ]^\top.
  \end{split}
\end{equation}
This frequency-independent processing is inspired by the conventional signal processing-based noise reduction algorithms, such as noise density estimation and wiener filter~\cite{Gannot2017ACP}.
It allows for a detailed analysis of the signal's local spectral pattern and stationarity. Experimental evidence supports the effective integration of a full-band model and a sub-band model within a single framework~\cite{hao_fullsubnet_2021}. However, the computational challenge for the existing full-band and sub-band modeling methods lies in their sub-band part, which processes each sub-band at the same granularity. This approach contrasts with the human auditory system, which is more sensitive to low-frequency sound and less sensitive to high-frequency sound~\cite{yost2007auditory}. Motivated by human auditory perception, 
we introduce a frequency partitioning technique that applies different processing granularity across the frequency bands to address this issue.

\subsubsection{Sub-Band Processing Based on Frequency Partitioning}
To account for the varying perceptual importance of different frequency bins, the input magnitude spectral is divided into $K$ non-overlapping frequency partitions:
\begin{equation}
  \mathcal{P} = \{ \mathcal{P}_1, \mathcal{P}_2, \dots, \mathcal{P}_K \} \quad \text{where} \quad \sum_{k=1}^{K} |\mathcal{P}_k| = F,
\end{equation}
where $|\cdot|$ is the size of the frequency partition set. Each frequency partition $\mathcal{P}_k$ is defined by a set of contiguous discrete frequency bins, thereby covering the entire frequency spectrum. The lowest frequency partition $\mathcal{P}_1$ includes the indices of the discrete frequency bins $\{1, \dots, f_{c_1}\}$, where $f_{c_1}$ is the cutoff frequency for this first partition. Each subsequent partition $\mathcal{P}_k$ for $k = 2, \dots, K-1$ starts from the cutoff frequency of the previous partition $f_{c_{k-1}}+1$ and extends up to its own cutoff frequency $f_{c_k}$. The final partition $\mathcal{P}_K$ contains frequencies from $f_{c_{K-1}}+1$ to the upper limit $F$. This partitioning can be formally expressed as
\begin{equation}
  \begin{split}
    \mathcal{P}_1 & = \{1, \dots, f_{c_1}\}, \\
    \mathcal{P}_k & = \{f_{c_{k-1}}+1, \dots, f_{c_k}\} \quad \text{for} \quad k = 2, \dots, K-1, \\
    \mathcal{P}_K & = \{f_{c_{K-1}}+1, \dots, F\}.
  \end{split}
\end{equation}
For each frequency partition $\mathcal{P}_k$, the Spiking-FullSubNet model adjusts the processing granularity by setting a grouping parameter $g_k$ that dictates how many discrete frequency bins to be processed jointly in the sub-band model. Specifically, for each frequency partition $\mathcal{P}_k$, we process groups of $g_k + 1$ adjacent discrete frequency bins, along with a context window of $N$ bins on either side, and the corresponding embedding vector output by the full-band model:
\begin{equation}
  \begin{split}
    \mathbf{x}_f^{k}(n) = & [ \underbrace{|x(n, f - N)|, \dots, |x(n, f-1)|}_\text{Lower neighboring frequencies}, \\
        & |x(n, f)|, |x(n, f+1)|, \dots, |x(n, f + g_k)|, \\
        & \mathbf{E}(n, f), \mathbf{E}(n, f + 1), \dots, \mathbf{E}(n, f + g_k), \\
        & \underbrace{|x(n, f + g_k + 1)|, \dots, |x(n, f + g_k + N)|}_\text{Higher neighboring frequencies} ]^\top,
  \end{split}
\end{equation}
where $\mathbf{x}_f^{k}(n)$ is the grouped feature vector for the $k^{th}$ frequency partition at time frame $n$, and $f$ is the starting frequency bin of the group within the partition $\mathcal{P}_k$. The value of $f$ ranges from the lower bound of $\mathcal{P}_k$ to an upper bound, ensuring the group of $g_k + 1$ bins is within the partition. The grouping parameter $g_k$ allows for a flexible adjustment of the sub-band resolution and computational complexity within each sub-band. For lower frequency partitions, where finer sub-band resolution is often more important for speech perception, $g_k$ may be set to a smaller value. Conversely, for higher frequency intervals, $g_k$ may be larger, reflecting the reduced perceptual importance of spectral details and allowing for reduced computational cost. This frequency partition processing strategy enables the Spiking-FullSubNet model to efficiently handle the spectral information with varying resolution across the frequency spectrum, which is more aligned with the non-uniform frequency resolution of human hearing.

\subsection{Learning Target}
Conventional speech enhancement methods work within the Short-Time Fourier Transform (STFT) domain~\cite{xu2013experimental,xu2014regression}. The typical methodology involves the estimation of time-frequency (TF) masks via neural networks, which are then applied element-wise to the complex STFT of the noisy speech mixture to extract the enhanced signal. 
However, the performance of this methodology will degrade if the frequency resolution gets too low~\cite{schroter_deepfilternet_2022}. Recently, multi-frame deep filtering (MFDF)~\cite{schroter_deepfilternet_2022} in the frequency domain has been proposed, where a filter is applied to multiple adjacent TF bins, enabling recovery of degraded signals
like notch-filters or time-frame zeroing.

The proposed Spiking-FullSubNet estimates a complex-valued 
MFDF filter for each TF bin within the STFT domain. We set the filter length (order) $d_k$ for input within the $k^{th}$ frequency partition and define the noisy multi-frame vector of the $\mathbf{x}_f^{k}(n)$ for the $k^{th}$ frequency partition as
\begin{equation}
  \begin{split}
    & \mathbf{\overline{x}}_{f}^{k}(n) = \\
    & \begin{bmatrix}
      x(n, f)       & x(n-1, f)         & \ldots & x(n - d_k, f)       \\
      x(n, f + 1)   & x(n-1,f + 1)      & \ldots & x(n - d_k, f + 1)   \\
      \vdots        & \vdots            & \ddots & \vdots              \\
      x(n, f + g_k) & x(n - 1, f + g_k) & \ldots & x(n - d_k, f + g_k)
    \end{bmatrix},
  \end{split}
\end{equation}
which includes only the streaming historical frames, thereby avoiding the introduction of future latency. For the noisy multi-frame input $\mathbf{x}_f^{k}(n)$, the output complex-valued MFDF filter is
\begin{equation}
  \begin{split}
    & \mathbf{\overline{w}}_{f}^{k} (n) = \\
    & \begin{bmatrix}
      w_0(f)       & w_1(f)       & \ldots & w_{d_k}(f)       \\
      w_0(f + 1)   & w_1(f + 1)   & \ldots & w_{d_k}(f + 1)   \\
      \vdots       & \vdots       & \ddots & \vdots           \\
      w_0(f + g_k) & w_1(f + g_k) & \ldots & w_{d_k}(f + g_k)
    \end{bmatrix}.
  \end{split}
\end{equation}
The application of the MFDF
is then expressed as
\begin{equation}
  \begin{split}
    \hat{\mathbf{s}}_f^{k}(n) & = \sum_{j} \left(\mathbf{\overline{w}}_{f}^{k} (n) \odot \mathbf{\overline{x}}_f^{k}(n) \right)_{ij} \\
    & = [ \hat{s}(n, f), \hat{s}(n, f + 1), \ldots \hat{s}(n, f + g_k) ]^\top,
  \end{split}
\end{equation}
where $\odot$ denotes the complex-valued hadamard product, $\hat{\mathbf{s}}_f^{k}(n)$ is the enhanced counterpart, and $\sum_{j} (\cdot)_{ij}$ means the sum by rows.
Our Spiking-FullSubNet model accommodates variable deep filtering orders across different frequency partitions. Lower frequencies, which exhibit stronger temporal correlations, are processed with a higher deep filtering order, allowing the network to capture more complex temporal structures. In contrast, higher frequencies characterized by weaker temporal correlations are processed with a lower deep filtering order. This approach promotes computational efficiency without sacrificing the resolution of critical spectral details, thus maintaining the integrity of the speech signal.


\subsection{Loss Function and SNN training}
Inspired by Braun \textit{et al.} \cite{Braun2020ACV}, we use a linear combination of magnitude loss $\mathcal{L}_{\text{Mag.}}$ and complex loss $\mathcal{L}_{\text{RI}}$ in the TF-domain:
\begin{equation}
  \begin{aligned}
    \mathcal{L}_{\text{TF}}    =& \alpha\, \mathcal{L}_{\text{Mag.}}+ (1-\alpha)\,\mathcal{L}_{\text{RI}}, \\
    \mathcal{L}_{\text{Mag.}}  = &\mathbb{E}_{s,\hat{s}} \big[\| s(n,f) - \hat{s}(n,f)  \|^2 \big],         \\
    \mathcal{L}_{\text{RI}}    =& \mathbb{E}_{s_r,\hat{s}_r} \big[\| s_r(n,f) - \hat{s}_r(n,f) \|^2\big]  \\
                              & + \mathbb{E}_{s_i,\hat{s}_i} \big[\| s_i(n,f) - \hat{s}_i(n,f) \|^2\big],
  \end{aligned}
  \label{TF loss}
\end{equation}
where $\alpha$ is a hyperparameter to balance the contribution from two losses. Moreover, an additional penalization on the waveform features is proven to help improve the speech quality \cite{Abdulatif2020InvestigatingCL}. In our model, we use the Scale-Invariant Signal-to-Distortion Ratio (SI-SDR)~\cite{LeRoux2018SDRH} loss function, which is computed directly in the time domain and forces the model to learn how to precisely estimate the magnitude and the phase of the target speech signals:
\begin{equation}
  \mathcal{L}_\text{SI-SDR} = - 10 \log_{10} \left( \frac{\| \frac{\hat{\mathbf{s}} ^T \mathbf{s}}{\|\mathbf{s}\|^2} \mathbf{s} \|^2}{\| \frac{\hat{\mathbf{s}}^T \mathbf{s}}{\|\mathbf{s}\|^2} \mathbf{s} - \hat{\mathbf{s}} \|^2} \right).
\end{equation}
The final loss function is formulated as follows:
\begin{equation}
  \mathcal{L} =\gamma_1 \,\mathcal{L}_{\text{TF}} + \gamma_2 \, (100 - \mathcal{L}_{\text{SI-SDR}}),
\end{equation}
where $\gamma_1$ and $\gamma_2$ are the weights of the corresponding losses.

With the established loss function, the Spiking-FullSubNet model can thus be trained end-to-end using backpropagation through time (BPTT)~\cite{werbos1990backpropagation}. However, BPTT cannot be directly used since the gradient of the spike generation function is zero almost everywhere except for the firing threshold, where it is infinity. To address this issue, we adopt a surrogate gradient to circumvent the non-differentiability issue~\cite{wu2018spatio,neftci2019surrogate}, formulated by
\begin{equation}
  \label{eq:triangle}
  \frac{\partial o_{i}^{l}(t)}{\partial u_{i}^{l}(t)}=\max \left(0,~1-|u_{i}^{l}(t)-\vartheta|\right).
\end{equation}

\section{Experimental Setup}
\label{sec:exp_setup}

In this section, we evaluate the proposed Spiking-FullSubNet on the publicly available speech enhancement dataset. Our study focuses on both the audio quality and energy efficiency metrics that are critical for edge devices.

\subsection{Intel N-DNS Challenge Dataset for Speech Enhancement}
We utilize the publicly available Intel N-DNS Challenge dataset~\cite{timcheck_intel_2023} for the speech enhancement performance evaluation. This comprehensive dataset contains a wide array of human speech audio samples in multiple languages, including English, German, French, Spanish, and Russian, as well as several noise categories.
The official Intel N-DNS Challenge repository provides a synthesizer script that generates clean (ground truth), noise (additive), and noisy (ground truth plus noise) audio segments for both training and validation. For testing, the challenge has officially released a large test set for a fair performance comparison.
With this official synthesizer script, we synthesize two subsets: a 495-hour subset for training and a 5-hour subset for validation. All audio samples, synthesized at a sampling rate of 16 kHz, are kept at a consistent duration of 30 seconds. For audios shorter than 30 seconds, we concatenate them with other speech signals from the same speaker, inserting a 0.2-second silence interval between clean speech utterances. The noisy audio is simulated using randomly selected speech and noise data, with SNRs ranging from -5 to 20 dB. We apply loudness normalization to each noisy audio sample to simulate agnostic input loudness levels from -35 to -15 decibels relative to full scale (dBFS).
For audio quality evaluation, we employ the metrics specified by the Intel N-DNS Challenge, which include SI-SDR~\cite{LeRoux2018SDRH}, and Deep Noise Suppression Mean Opinion Score (DNSMOS)~\cite{Reddy2020DnsmosAN}. We also employ the power consumption metrics that will be introduced in Section~\ref{sec:power-consumption}.

\subsection{Power Consumption Measurement}
\label{sec:power-consumption}
The power consumption of a neuromorphic system is roughly proportional to the total number of computational primitives used, including synaptic operations (\text{SynOPs}) and neuron operations (\text{NeuronOPs})~\cite{davies2018loihi,timcheck_intel_2023}. 
Based on the power estimation conducted on the Intel Loihi architecture, which indicates that the energy consumed by one NeuronOP is approximately equivalent to that of about $10 \times \text{SynOPs}$~\cite{timcheck_intel_2023}, we derive the power consumption proxy $P_{\text{proxy}}$ using the following formula:
\begin{equation}
  P_{\text{proxy}} = \text{SynOPs} + 10 \times \text{NeuronOPs},
\end{equation}
\begin{equation}
  \text{SynOPs} = \sum_{l=1}^{L-1} \sum_{i=1}^{\mathcal{N}^{l}}\mathcal{R}_i^{l} (\mathcal{N}^{l+1} + \mathcal{N}^{l}),
  \label{synops}
\end{equation}
\begin{equation}
  \text{NeuronOPs} = \sum_{l=1}^L \mathcal{N}^{l},
\end{equation}
where $\mathcal{R}_i^l$ denotes the firing rate of neuron $i$ in layer $l$, $\mathcal{N}^{l}$~represents the number of neurons in layer $l$, and $L$ is the total number of layers in the network. Eq. (\ref{synops}) is formulated based on the recurrent neural network architecture, encompassing both feedforward ($\mathcal{N}^{l+1}$) and recurrent outputs ($\mathcal{N}^{l}$). 

We also utilize the power delay product (PDP) metric~\cite{davies2018loihi,timcheck_intel_2023}, which consolidates latency and power consumption into a single metric, to assess and compare different systems with distinct focuses on speed and power efficiency. The PDP proxy~$\text{PDP}_{\text{proxy}}$ is defined by
\begin{equation}
  \text{PDP}_{\text{proxy}} = P_{\text{proxy}} \times \text{Latency}.
\end{equation}

For the computational costs of Multiply-ACcumulate (MAC) and ACcumulate (AC) operations employed for \text{SynOPs} and \text{NeuronOPs} respectively, we refer to the findings presented in~\cite{han2015learning}. These findings indicate that with $45nm$ CMOS technology, one floating-point MAC operation consumes $4.6pJ$ of energy while one AC operation consumes $0.9~pJ$.


\subsection{Implementation Details}

All speech audio data are processed at a sampling rate of 16 kHz. The STFT is set up with a window length of 32 ms (512 samples) and a hop length of 8 ms (128 samples), utilizing a Hanning window and comprising 512 FFT frequency bins. The proposed Spiking-FullSubNet takes the magnitude spectrogram as its input. We utilize the AdamW optimizer~\cite{Loshchilov2017DecoupledWD} with a learning rate of $1 \times 10^{-3}$ and set the gradient norm clipping to 10. In the loss function $\mathcal{L}$, the weights of different terms are set to $\alpha = 0.5, \gamma_1=0.5, \gamma_2=0.001$. We set the number of neighboring frequency bins to 15. To further improve energy efficiency, we add the total number of synaptic operations into the loss function to penalize excessive firing.
We develop different variants of the Spiking-FullSubNet with varying model sizes, detailed in Table \ref{tab:para_ablation}. These variants differ in aspects, including the granularity of frequency partitioning and the order of deep filtering. 

We divide all discrete frequency bins based on the study of human auditory perception~\cite{sataloff1992human,sataloff1992human} to see the effects of frequency partitioning configuration on denoising performance and model efficiency.
The fundamental frequency, a crucial characteristic of speech signals, extends up to 1 kHz. 
Therefore, we group the lowest 32 discrete frequency bins, corresponding to $0 \sim 1$ kHz, applying the smallest grouping size within this range. 
Then, given that the human voice's most significant content lies between $1 \sim 4$ kHz, a range particularly sensitive for the human auditory system and rich in harmonic structures, we group the following 96 discrete frequency bins, equivalent to $1 \sim 4$ kHz, with uniform processing granularity in this range. 
Lastly, we group the remaining high-frequency bins, 128 discrete frequency bins corresponding to $4 \sim 8$ kHz, as a separate group. These high-frequency bins have a more minor impact on human auditory perception, allowing for coarser granularity in processing. In summary, we establish three frequency partitions: $0 \sim 1$ kHz, $1 \sim 4$ kHz, and $4 \sim 8$ kHz, each tailored to optimize computational efficiency and performance based on human auditory characteristics. Additional implementation details can be found in our open-source code repository.

\begin{table*}[!t]
	\centering
	\caption{Comparison of speech enhancement performance among various baseline models on the Intel N-DNS Challenge dataset. The evaluation metrics include SI-SNR, SI-SNRi, DNSMOS, latency, and computational efficiency. The table comprises the official baselines from the Intel N-DNS Challenge, state-of-the-art real-time ANN models, and top-performing SNN models from the competition. Results are presented with the year of publication or the corresponding challenge rankings.}
	\label{tab:intelndns}
 \resizebox{\textwidth}{!}{%
	\renewcommand{\arraystretch}{1.2}
	\addtolength{\tabcolsep}{-0.2em}
	\begin{tabular}{cccccccccccccc}
		\toprule
		\multirow{3}{*}{\textbf{Entry}}                      & \multirow{3}{*}{\begin{tabular}[c]{@{}c@{}}\textbf{Year} / \\ \textbf{(Rank)}\end{tabular}} & \multirow{3}{*}{\begin{tabular}[c]{@{}c@{}}\textbf{SI-SNR ($\uparrow$)} \\ \textbf{(dB)}\end{tabular}} & \multicolumn{2}{c}{\textbf{SI-SNRi ($\uparrow$)}} & \multicolumn{3}{c}{\textbf{DNSMOS ($\uparrow$)}} & \multicolumn{2}{c}{\textbf{Latency ($\downarrow$)}} & \multirow{2}{*}{\begin{tabular}[c]{@{}c@{}}\textbf{Power}\\ \textbf{Proxy}~($\downarrow$) \end{tabular}} & \multirow{2}{*}{\begin{tabular}[c]{@{}c@{}}\textbf{PDP} \\ \textbf{Proxy} ($\downarrow$)\end{tabular}} & \multirow{2}{*}{\begin{tabular}[c]{@{}c@{}}\textbf{Energy} \\ \textbf{Cost} ($\downarrow$)\end{tabular}} & \multirow{2}{*}{\begin{tabular}[c]{@{}c@{}}\textbf{Param}\\\textbf{Count} \end{tabular}}                                                                               \\
		\cmidrule{4-10}
		                                                     &                 &       & \textbf{data} & \textbf{enc+dec} & \multirow{2}{*}{\textbf{OVR}} & \multirow{2}{*}{\textbf{SIG}       } & \multirow{2}{*}{\textbf{BAK}} & \textbf{enc+dec} & \textbf{total} &                                         &                                    &                                    &                                              \\
		                                                     &                 &       & \textbf{(dB)} & \textbf{(dB)}    &                               &                                      &                               & \textbf{(ms)}    & \textbf{(ms)}  & \multicolumn{1}{c}{\textbf{(Ops/s)}}    & \multicolumn{1}{c}{\textbf{(Ops)}} & \multicolumn{1}{c}{\textbf{($J$)}} & \multicolumn{1}{c}{\textbf{($\times 10^3$)}} \\
		\midrule
		Noisy (Unproc.)                                      & -               & 7.37  & -             & -                & 2.43                          & 3.16                                 & 2.69                          & -                & -              & -                                       & -                                  & -                                  & -                                            \\
		\midrule
		\multicolumn{14}{l}{\textcolor[rgb]{0.502,0.502,0.502}{\textit{Official Intel N-DNS Challenge baseline systems. The results are directly quoted from the Intel N-DNS official repository\footnotemark.}}}                                                                                                                                                                                                                                                                                                                                                                                                                                                                                                                                                                                                                                                                                                          \\
		Microsoft NsNet2~\cite{dubey2022icassp}              & 2023            & 11.63 & 4.26          & 4.26             & 2.95                          & 3.26                                 & 3.93                          & 0.02             & 20.02          & 136.13 M                                & 2.72 M                             & 12.51 $\mu$                        & 2681                                         \\
		Intel DNS Network~\cite{timcheck_intel_2023}         & 2023            & 12.51 & 5.14          & 5.14             & 3.08                          & 3.34                                 & 4.07                          & 0.02             & 32.02          & -                                       & -                                  & -                                  & 1901                                         \\
		SDNN Network~\cite{timcheck_intel_2023}              & 2023            & 12.26 & 4.89          & 4.89             & 2.70                          & 3.20                                 & 3.45                          & 0.02             & 32.02          & 14.52 M                                 & 0.46 M                             & 0.41 $\mu$                         & 525                                          \\
		\midrule
		\multicolumn{14}{l}{\textcolor[rgb]{0.502,0.502,0.502}{\textit{State-of-the-art real-time ANN baselines, which are adapted from their official code repositories and trained on the Intel N-DNS Challenge dataset.}}}                                                                                                                                                                                                                                                                                                                                                                                                                                                                                                                                                                                                                                                                                              \\
		DCCRN\footnotemark~\cite{hu_dccrn_2020}              & 2021            & 12.46 & 5.09          & 5.09             & 2.85                          & 3.11                                 & 3.77                          & 0.02             & 20.02          & 5.07 G                                  & 0.10 G                             & 0.46 $m$                           & 1247                                         \\
		FullSubNet\footnotemark~\cite{hao_fullsubnet_2021}   & 2022            & 13.55 & 6.18          & 6.18             & 2.93                          & 3.22                                 & 3.84                          & 0.03             & 32.03          & 3.65 G                                  & 0.12 G                             & 0.55 $m$                           & 1141                                         \\
		Fast FullSubNet\footnotemark[3]~\cite{hao_fast_2023} & 2023            & 13.88 & 6.51          & 6.51             & 2.91                          & 3.24                                 & 3.82                          & 0.03             & 32.03          & 0.49 G                                  & 0.02 G                             & 0.09 $m$                           & 1141                                         \\
		CMGAN\footnotemark~\cite{cao_cmgan_2024}             & 2024            & 14.01 & 6.64          & 6.64             & 2.81                          & 3.23                                 & 3.84                          & 0.02             & 20.02          & 15.94 G                                 & 0.32 G                             & 1.47 $m$                           & 1413                                         \\
		\midrule
		\multicolumn{14}{l}{\textcolor[rgb]{0.502,0.502,0.502}{\textit{Intel N-DNS Challenge top-ranking systems. The results are directly quoted from the Intel N-DNS official repository\footnotemark[1].}}}                                                                                                                                                                                                                                                                                                                                                                                                                                                                                                                                                                                                                                                                                                             \\
		CTDNN LAVADL\footnotemark[1]                         & rank 2          & 13.52 & 6.59          & 6.59             & 2.97                          & 3.32                                 & 3.86                          & 0.00             & 32.00          & 61.37 M                                 & 1.96 M                             & 1.76 $\mu$                         & 905                                          \\
		Sparsity SDNN\footnotemark[1]                        & rank 3          & 12.16 & 4.80          & 4.80             & 2.70                          & 3.19                                 & 3.46                          & 0.01             & 32.01          & 9.32 M                                  & 0.30 M                             & 0.27 $\mu$                         & 344                                          \\
		PSNN\footnotemark[1]                                 & rank 4          & 12.32 & 4.96          & 4.96             & 2.68                          & 2.91                                 & 3.96                          & 0.00             & 32.00          & 57.24 M                                 & 1.83 M                             & 1.65 $\mu$                         & 724                                          \\
		\midrule
		\textbf{Spiking-FullSubNet}                          & \textbf{winner} & 15.20 & 7.83          & 7.83             & 3.03                          & 3.35                                 & 3.94                          & 0.02             & 32.02          & 51.30 M                                 & 1.64 M                             & 1.48 $\mu$                         & 965                                          \\
		\bottomrule
	\end{tabular}
 }
\end{table*}

\section{Results and Discussion}
\label{sec:results}

In this section, we first evaluate the speech enhancement performance of the proposed Spiking-FullSubNet. Subsequently, we conduct ablation studies to analyze the impact of different components within the Spiking-FullSubNet as well as their configurations. Finally, we analyze the temporal information processing capability and energy efficiency of the proposed GSN neuron model. 

\subsection{Superior Speech Denoising Capability}

To ensure a comprehensive evaluation of the speech enhancement performance, this analysis employs a range of metrics concerning audio quality and power consumption. As summarized in Table~\ref{tab:intelndns}, we compare the proposed Spiking-FullSubNet with SOTA real-time ANN baselines and top-ranking SNN models from the Intel N-DNS Challenge. For a fair comparison, all models utilize the same STFT encoding and iSTFT decoding. Additionally, in order to provide a clear understanding of the characteristics of the noisy input audio, we also present the results for unprocessed noisy audio, indicated in the row labeled ``Noisy (Unproc.)".

\textit{Microsoft NsNet2}~\cite{dubey2022icassp} is the official baseline benchmark for the Microsoft DNS Challenge 2022, which consists of a series of LSTM layers optimized using a perception-based loss function.
\textit{Intel DNS Network}~\cite{timcheck_intel_2023} is a proprietary production-level model used in Intel's product environments, which is a streaming, real-time model that integrates LSTM and CNN. It is trained on a larger-scale dataset with data augmentation techniques. As claimed by the Intel N-DNS Challenge organizer~\cite{timcheck_intel_2023}: ``The network was trained using proprietary datasets and augmentation techniques, and as such we view its audio quality results as upper-bound aspirational
targets for challenge submissions." 
\textit{SDNN network}~\cite{timcheck_intel_2023} utilizes a sigma-delta method and is the official SNN baseline for the Intel N-DNS Challenge. From these results, we find our Spiking-FullSubNet model performs considerably better than the above models, demonstrating remarkable effectiveness.

\footnotetext[1]{\href{https://github.com/IntelLabs/IntelNeuromorphicDNSChallenge}{https://github.com/IntelLabs/IntelNeuromorphicDNSChallenge}}
\footnotetext[2]{\href{https://github.com/huyanxin/DeepComplexCRN}{https://github.com/huyanxin/DeepComplexCRN}}
\footnotetext[3]{\href{https://github.com/Audio-WestlakeU/FullSubNet}{https://github.com/Audio-WestlakeU/FullSubNet}}
\footnotetext[4]{\href{https://github.com/ruizhecao96/CMGAN}{https://github.com/ruizhecao96/CMGAN}}

We further compare our method against SOTA real-time ANN models. 
Among the models under consideration, \textit{DCCRN}~\cite{hu_dccrn_2020} adopts a complex-valued convolutional neural network and ranked 1st in the Microsoft DNS Challenge.
Both \textit{FullSubNet}~\cite{hao_fullsubnet_2021} and its variant \textit{Fast FullSubNet}~\cite{hao_fast_2023} utilize full-band and sub-band modeling techniques similar to our approach. Fast FullSubNet differs from FullSubNet by strategically using mel-scale inputs, which reduces the number of sub-band features.
However, our method distinguishes itself from these two models by employing a brain-inspired sub-band partitioning method and leveraging spiking neural networks, setting it apart from the aforementioned models.
Furthermore, we include the recently proposed SOTA \textit{CMGAN}~\cite{cao_cmgan_2024}, a complex-valued multi-scale generative adversarial network, in our comparison.
To ensure a fair comparison, we retrain the SOTA models on the Intel N-DNS Challenge dataset with a comparable number of parameters. Our modifications only involve adjusting the number of hidden units in RNNs or the output channels in CNNs, without altering their core architectures.
In comparison to these SOTA ANN models, it is evident that our proposed model, Spiking-FullSubNet, outperforms them all in terms of both audio quality and energy efficiency. Notably, when compared with the best-performing ANN model \textit{CMGAN}, the energy consumption of our Spiking-FullSubNet model is almost three orders of magnitude smaller (1.47 $m$ vs. 1.48 $\mu$). This significant energy efficiency is attributable to the cheaper and sparser operations inherent to the SNN model.

In addition, we also compare our method with top-ranking systems~\cite{timcheck_intel_2023} in the Intel N-DNS Challenge, as shown in the table. 
While the repositories for these methods are anonymous, preventing a direct comparison of specific implementations, it's important to note that all submissions to the Intel N-DNS Challenge underwent rigorous verification by the organizers. This inspection ensures the authenticity of their model's real-time capabilities, power consumption, and overall performance. Our Spiking-FullSubNet model stands out from these competitors and has won the championship for the algorithmic track of the Intel N-DNS Challenge. Remarkably, it achieves the highest DNSMOS overall score and SI-SNR, suggesting superior denoising capability. Additionally, the power proxy and PDP proxy are 51.30 M-Ops/s and 1.64 M-Ops, respectively, indicating a good balance between audio enhancement performance and computational efficiency. 

\begin{figure}[!t]
    \centering
    \includegraphics[width=\linewidth]{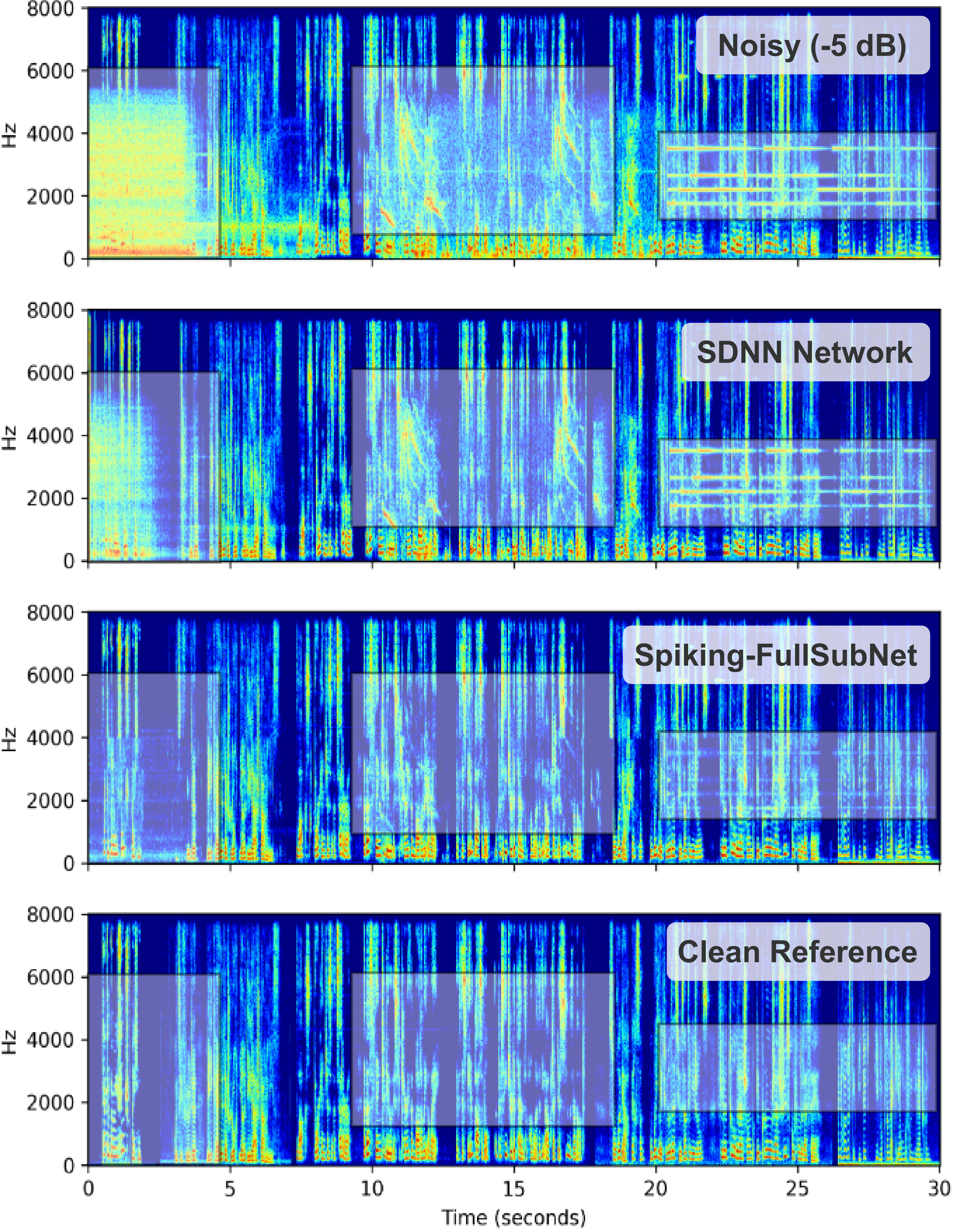}
    \caption{Comparison of time-frequency magnitude spectrograms for an example from the Intel N-DNS Challenge test set.}
    \label{fig:visual_enh_sample}
\end{figure}

Finally, in Figure~\ref{fig:visual_enh_sample}, we provide a detailed visual comparison of the spectrograms for a single test sample, showcasing the noisy input signal, enhanced signal, and the clean reference signal. The noisy speech input has an SNR of -5 dB and contains both stationary and non-stationary noise components, presenting a significant challenge for speech enhancement systems. It can be seen that the Spiking-FullSubNet model not only preserves essential speech features but also significantly reduces the background noise. In contrast, the Intel N-DNS official baseline model, referred to as the ``SDNN network", struggles to handle the various noise components, as highlighted by the residual artifacts and noise components within the white translucent rectangles. This comparison underscores the effectiveness of our model in addressing the challenges posed by diverse noise conditions and illustrates its potential for practical implementations in real-world speech enhancement scenarios.

\subsection{Ablation Studies for Core Components of Spiking-FullSubNet}
To further understand the effectiveness of different components within the proposed Spiking-FullSubNet, we conduct ablation studies by replacing the proposed GSN model with other commonly used spiking neuron models, removing the sub-band modeling, and removing deep filtering components from Spiking-FullSubNet.
As shown in Table~\ref{tab:comp_ablation}, our model equipped with the newly proposed GSN model demonstrates superior performance with an SI-SNR of 15.20 dB and DNSMOS scores of 3.03 (OVR), 3.35 (SIG), and 3.94 (BAK). When the GSN is replaced with the LIF neuron, there is a noticeable drop in performance, with the SI-SNR falling to 9.72 dB. This decline is also visible in the DNSMOS metrics, which decreases across all categories, and most notably in the overall rating (OVR), which drops to 2.73. The use of the PLIF neuron, while yielding better results than LIF with an SI-SNR of 11.37 dB, still underperforms our GSN model by a large margin. This decrease is consistent across the DNSMOS metrics.
The performance of ALIF is better than that of PLIF, but the margin is limited. These results indicate that the proposed GSN model is favorably advantageous in the speech enhancement task, outperforming existing spiking neuron models.

In addition, we remove the sub-band models from Spiking-FullSubNet and only keep the prior full band model to verify the improvement brought by the sub-band models. To make a fair comparison, we further increase the network layers and parameter counts
of the remaining full band model. From the table, we can notice that removing the sub-band components from Spiking-FullSubNet results in decreased performance, with the SI-SNR dropping to 13.97 dB and DNSMOS scores decreasing across all categories. Our result indicates that the sub-band component can effectively learn to focus on complementary cues to the full-band component, leading to better speech enhancement performance. This observation aligns with findings from other full-band and sub-band modeling studies~\cite{hao_fullsubnet_2021,wang_tf-gridnet_2023}.

Finally, we observe that removing the deep filtering component from the Spiking-FullSubNet will also result in lower performance. The SI-SNR drops to 15.10 dB, and DNSMOS scores decrease across all categories. This suggests that the deep filtering component enhances the temporal information processing capability of the Spiking-FullSubNet, further contributing to its improved speech enhancement performance.


\begin{table}[!t]
  \centering
  \caption{Ablation Studies for Core Components of Spiking-FullSubNet. ``w/o Sub-band models" means the sub-band GSN models are removed from the Spiking-FullSubNet. To fairly compare the performance, ``w/o Sub-band models'' has more layers and parameters than that of the Spiking-FullSubNet.  ``w/o Deep Filtering'' means the deep filtering is fully removed from the Spiking-FullSubNet.}
  \label{tab:comp_ablation}
\renewcommand{\arraystretch}{1.2}
  \begin{tabular}{cccccc} 
\toprule
\multirow{2}{*}{\textbf{Entry}} & \multirow{2}{*}{\begin{tabular}[c]{@{}c@{}}\textbf{\# Para.} \\ ($\times 10^3$)\end{tabular}} & \multirow{2}{*}{\begin{tabular}[c]{@{}c@{}}\textbf{SI-SNR} \\ (dB)\end{tabular}} & \multicolumn{3}{c}{\textbf{DNSMOS}} \\ 
\cmidrule{4-6}
 &  &  & OVR & SIG & BAK \\ 
\midrule
Noisy & - & 7.37 & 2.43 & 3.16 & 2.69 \\ 
\midrule
Spiking-FullSubNet & 965 & 15.20 & 3.03 & 3.35 & 3.94 \\ 
\midrule
\multicolumn{6}{l}{\textcolor[rgb]{0.502,0.502,0.502}{\textit{Replace the GSN model with other spiking neuron models.}}} \\
\textit{w/} LIF neuron & 948 & 9.72 & 2.73 & 3.26 & 3.40 \\
\textit{w/} PLIF neuron & 950 & 11.37 & 2.82 & 3.25 & 3.65 \\
\textit{w/} ALIF neuron & 952 & 11.43 & 2.85 & 3.26 & 3.68 \\ 
\midrule
\multicolumn{6}{l}{\textcolor[rgb]{0.502,0.502,0.502}{\textit{Remove sub-band components 
(just a conventional full-band model). }}} \\
\textit{w/o} Sub-band models & 991 & 13.97 & 2.91 & 3.26 & 3.84 \\ \midrule
\multicolumn{6}{l}{\textcolor[rgb]{0.502,0.502,0.502}{\textit{Remove deep filtering.}}} \\
\textit{w/o} Deep Filtering & 918 & 15.10 & 3.01 & 3.28 & 3.94 \\
\bottomrule
\end{tabular}
\end{table}

\begin{table*}[]
\centering
\caption{The impact of different network components configurations within the Spiking-FullSubNet. $\{ g_k \}$ is the grouping parameter set. The value in $\{ g_k \}$ means how many discrete frequencies are processed jointly in each frequency partition. $\{ d_k \}$ is the deep filtering order set for all frequency partitions.}
\label{tab:para_ablation}
\resizebox{\textwidth}{!}{%
\begin{tabular}{@{}c|c|c|c|c|c|ccc|cccc@{}}
\toprule
\multicolumn{3}{c|}{\begin{tabular}[c]{@{}c@{}}Grouping Parameters \\ $\{ g_k \}$\end{tabular}} & \multicolumn{3}{c|}{\begin{tabular}[c]{@{}c@{}}Filter Orders \\ $\{ d_k \}$\end{tabular}} & \multirow{2}{*}{\begin{tabular}[c]{@{}c@{}}Param Count \\ ($\times 10^3$)\end{tabular}} & \multirow{2}{*}{\begin{tabular}[c]{@{}c@{}}Power Proxy \\ (M-Ops/s)\end{tabular}} & \multirow{2}{*}{\begin{tabular}[c]{@{}c@{}}PDP Proxy \\ (M-Ops)\end{tabular}} & \multirow{2}{*}{\begin{tabular}[c]{@{}c@{}}SI-SNR\\ (dB)\end{tabular}} & \multicolumn{3}{c}{DNSMOS} \\ \cmidrule(lr){1-3} \cmidrule(lr){4-6} \cmidrule(lr){11-13} 
0$\sim$1 KHz & 1$\sim$4 KHz & 4$\sim$8 KHz & 0$\sim$1 KHz & 1$\sim$4 KHz & 4$\sim$ 8 KHz & & & & & OVR & SIG & BAK \\ \midrule
\multicolumn{6}{c}{\textit{Unprocessed noisy speech}} & - & - & - & 7.37 & 2.44 & 3.16 & 2.69 \\ \midrule
1 & 1 & 1 & \multirow{8}{*}{3} & \multirow{8}{*}{1} & \multirow{8}{*}{1} & 832 & 1900.47 & 63.75 & 15.26 & 3.05 & 3.36 & 3.96 \\
2 & 32 & 64 & & & & 918 & 175.90 & 5.63 & 15.27 & 3.04 & 3.35 & 3.97 \\
4 & 32 & 64 & & & & 922 & 121.37 & 3.89 & 15.24 & 3.04 & 3.35 & 3.96 \\
8 & 32 & 64 & & & & 929 & 93.83 & 3.01 & 15.16 & 3.02 & 3.34 & 3.93 \\
16 & 32 & 64 & & & & 944 & 80.52 & 2.58 & 14.86 & 2.97 & 3.29 & 3.88 \\
32 & 32 & 64 & & & & 972 & 75.45 & 2.41 & 13.73 & 2.87 & 3.26 & 3.74 \\
8 & 96 & 64 & & & & 987 & 82.34 & 2.64 & 14.74 & 2.98 & 3.32 & 3.88 \\
8 & 32 & 128 & & & & 987 & 88.28 & 2.83 & 14.97 & 2.99 & 3.33 & 3.88 \\ \midrule
\multirow{6}{*}{8} & \multirow{6}{*}{32} & \multirow{6}{*}{64} & 1 & 1 & 1 & 922 & 93.21 & 2.98 & 15.10 & 3.01 & 3.28 & 3.94 \\
& & & 5 & 1 & 1 & 936 & 96.08 & 3.08 & 15.18 & 3.02 & 3.35 & 3.93 \\
& & & 7 & 1 & 1 & 943 & 97.46 & 3.12 & 15.11 & 3.00 & 3.21 & 3.87 \\
& & & 5 & 3 & 1 & 965 & 99.21 & 3.17 & 15.20 & 3.03 & 3.35 & 3.94 \\
& & & 5 & 5 & 1 & 994 & 103.72 & 3.32 & 15.11 & 3.03 & 3.33 & 3.96 \\
& & & 5 & 3 & 3 & 1023 & 105.44 & 3.37 & 15.02 & 2.97 & 3.21 & 3.77 \\ \bottomrule
\end{tabular}%
}
\end{table*}

\subsection{Ablation Studies for Different Network Configurations of Spiking-FullSubNet}
\label{sec:network_config}

In this section, we delve deeper into the impact of different network component configurations within Spiking-FullSubNet and present the results in Table~\ref{tab:para_ablation}. Specifically, we investigate the effects of varying the grouping parameter set ($\{ g_k \}$) and the deep filtering order set ($\{ d_k \}$) across all frequency partitions. To clearly demonstrate how these configurations influence power consumption, we remove the SynOPs regularization term during network training. In the following, we discuss how different configurations of $\{ g_k \}$ and $\{ d_k \}$ affect network performance.

\subsubsection{Grouping Parameters $\{ g_k \}$}
Processing each discrete frequency independently, by setting the grouping parameter to 1 for all partitions (i.e., $\{1, 1, 1 \}$, similar to the original FullSubNet~\cite{hao_fullsubnet_2021}), results in notably high performance. However, this approach incurs a very high computational cost of $1900.47~\text{M-Ops/s}$. By employing our proposed frequency partitioning method, which divides discrete frequencies according to human auditory perception characteristics and assigns different processing granularities to each partition, we significantly reduce computational cost while maintaining performance close to that achieved without frequency partitioning.

To be specific, we first investigate the impact of varying processing granularity for low frequencies (0 $\sim$ 1 kHz), ranging from fine to coarse (2 $\rightarrow$ 4 $\rightarrow$ 8 $\rightarrow$ 16 $\rightarrow$ 32). As expected, computational cost decreases as processing granularity becomes coarser, as the model processes more discrete frequencies simultaneously, reducing the required computational operations. In addition, we notice that initially increasing granularity has minimal impact on performance. However, beyond a certain point, performance degrades.
Specifically, setting the grouping parameter to 4 reduces the computational cost to 6.39\% of the baseline setting (i.e., $\{1,1,1\}$) while maintaining similar performance.
Setting the grouping parameter to 8 slightly decreases performance but reduces the computational cost to 4.94\% of the baseline. However, beyond this point, performance significantly degrades with grouping parameters set to 16 and 32. This degradation is likely due to overly coarse granularity hindering the model's ability to differentiate temporal stationarity between speech and noise signals in the frequency domain, which is crucial for effective sub-band processing.

Then, we investigate the impact of varying processing granularity for the mid-frequency partition (1 $\sim$ 4 kHz) while keeping the granularity for the low and high-frequency partitions constant. We observe a significant performance drop when increasing the granularity from 32 to 96 in the mid-frequency partition. This is likely because the mid-frequency partition contains many harmonic structures and exhibits distinct differences in the stationarity of speech and noise, similar to the low-frequency partition. Fine granularity is necessary to capture these details. Finally, we explore increasing processing granularity only for the high-frequency partition. While performance still declines, the decrease is less pronounced compared to the drop observed when coarsening the mid-frequency processing granularity. This can be explained by the fact that high-frequency components may possess more redundant information for human auditory perception, allowing for coarser granularity without a significant loss in model performance.

In summary, varying the grouping parameters within different frequency partitions can substantially reduce the computational burden for sub-band processing, demonstrating the effectiveness of our approach. However, while computationally efficient, excessively coarse grouping parameters tend to decrease performance. To achieve a balance between performance and computational cost, we will use $\{8, 32, 64\}$ as our default configuration for subsequent ablation studies.

\subsubsection{Deep Filter Orders $\{ d_k \}$}
We further investigate the impact of deep filtering on Spiking-FullSubNet by varying the order of the deep filtering for each frequency partition. 
As shown in Table~\ref{tab:para_ablation}, we begin by setting the order to $\{1, 1, 1\}$, meaning deep filtering is not utilized as a training target. When increasing the deep filtering order for the $0 \sim 1$ kHz partition gradually improves performance as the order increases from 1 to 3 and then to 5. However, further increasing the order to 7 results in a performance decline. This suggests that while higher-order deep filtering is beneficial in the low-frequency partition, an excessively high order introduces too much redundant historical information. This prevents effective utilization of short-time correlations within the speech signal, which is the fundamental rationale behind the deep filtering~\cite{schroter_deepfilternet_2022}.
Similarly, for the mid-frequency partition, increasing the filter order continues to be beneficial. However, when the order reaches 5, a slight decrease in SI-SNR is observed, i.e., from 15.20 to 15.11. 
However, in the high-frequency partition, applying deep filtering negatively impacts performance. Even a small order of 3 results in a noticeable performance drop. This may be because short-term correlations are less prevalent at high frequencies.

\subsection{Enhanced multi-scale Temporal Information Processing}

\begin{figure}[!tb]
  \centering\includegraphics[width=0.47\textwidth]{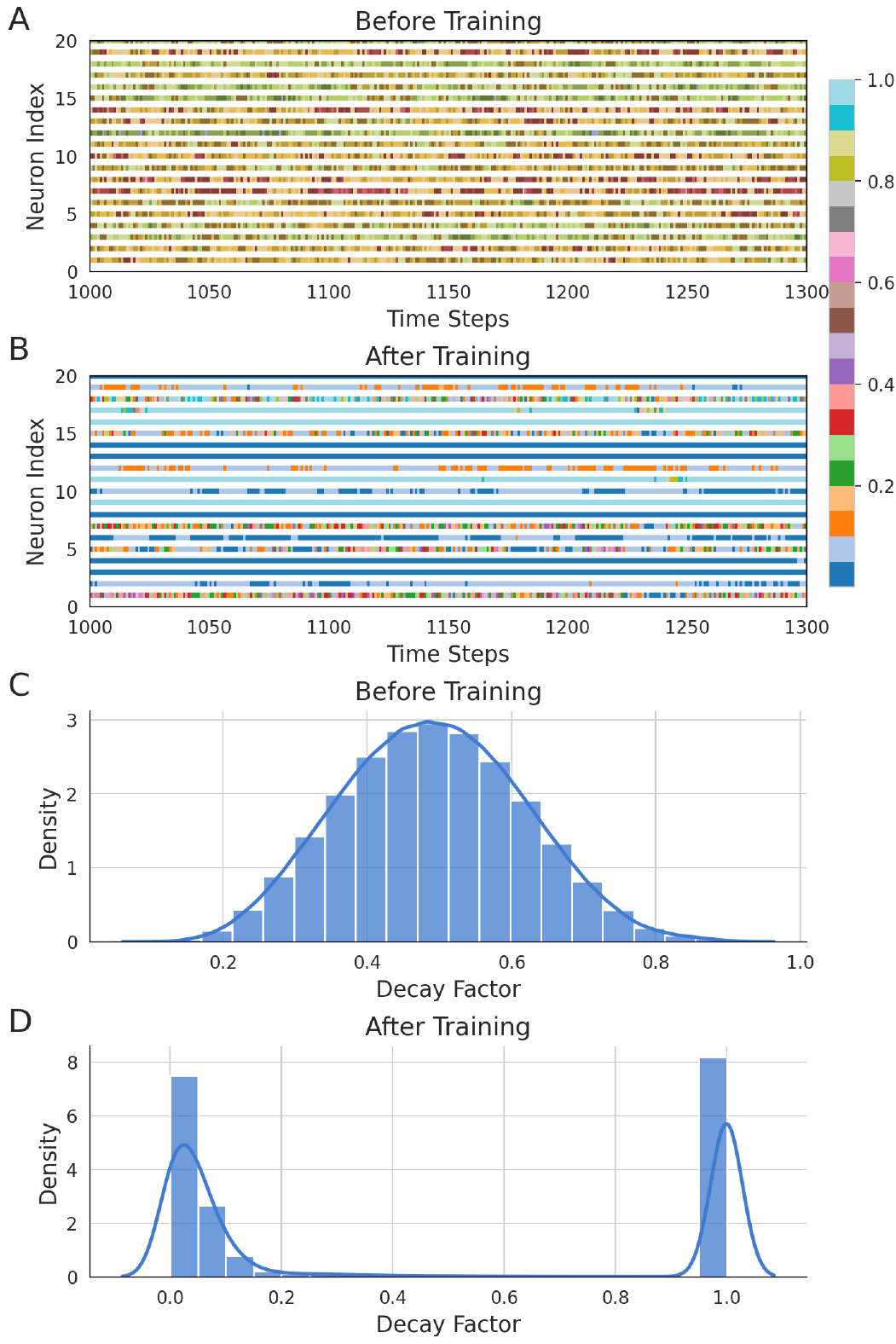}
  \caption{Analyze the multi-scale temporal information processing capability of GSN. \textbf{(A)} and \textbf{(B)} depicts the evolution of decay factors over time, before and after training, respectively. \textbf{(C)} and \textbf{(D)} present the distributions of decay factors across all neurons and time steps, again comparing the states before and after the training process.}
  \label{Fig:hist_gate}
\end{figure}

Unlike existing spiking neuron models, our GSN model can dynamically adjust decay factors over time in response to inputs. This dynamic adjustment of decay factors facilitates multi-scale temporal information processing, which is critical for achieving high performance in speech denoising tasks. To demonstrate this capability, we visualize the evolution of decay factors over time, as shown in Figs.~\ref{Fig:hist_gate}\textbf{A} and \ref{Fig:hist_gate}\textbf{B}. Initially, the decay factors are relatively uniform across different time steps. However, after training, the decay factors show strong temporal variations, aligning with the temporal dynamics of the input signals. This suggests that the GSN model is able to adaptively modulate its temporal processing characteristics to better match the characteristics of the input speech signals. We further investigate the distributions of decay factors across all neurons and time steps. As shown in Figs.~\ref{Fig:hist_gate}\textbf{C} and \ref{Fig:hist_gate}\textbf{D}, the initial decay factors closely follow a Gaussian distribution, with a mean value of approximately 0.5. However, after training, the decay factors exhibit a bimodal distribution, focusing on both slow (decay value close to 1) and rapid decay (decay value close to 0). This distribution shift allows the GSN model to effectively process multiple sound sources that have different temporal dynamics, which is a key requirement for successful speech denoising. The dynamic and adaptive nature of the decay factors in the GSN model is a significant advancement over existing spiking neuron models, enabling more effective multi-scale temporal processing for speech denoising and potentially other time-series signal processing tasks.

\subsection{High Energy Efficiency}
\begin{figure}[!tb]
  \centering\includegraphics[width=0.47\textwidth]{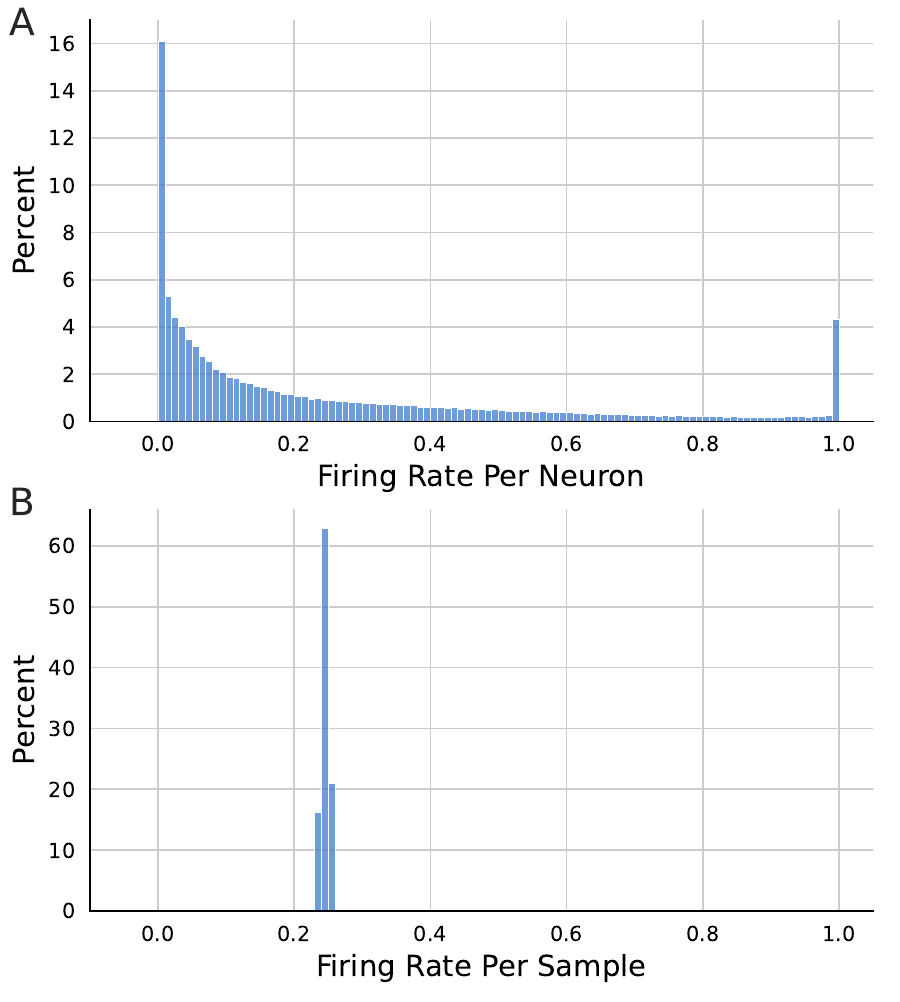}
  \caption{(\textbf{A}) Distributions of per neuron's average firing rate, and (\textbf{B}) Distributions of per sample's average firing rate.
  }
  \label{Fig:hist_fr}
\end{figure}

To assess the energy efficiency of our proposed Spiking-FullSubNet architecture, we consider the sparsity of output spikes, which is directly related to the computational cost. Specifically, we evaluate the firing rates of the neurons within the trained Spiking-FullSubNet on the test set of the Intel N-DNS dataset~\cite{timcheck_intel_2023}. As depicted in Fig.~\ref{Fig:hist_fr}\textbf{A}, the results show a high degree of sparsity in the neuronal activity. Over 16\% of neurons remain inactive throughout the testing phase, and approximately 60\% of neurons exhibit a firing rate of less than 0.2. This level of sparsity in neuronal activity is beneficial for achieving high energy efficiency on event-driven neuromorphic hardware implementations.
Next, we calculate the firing rate for each sample by averaging the firing rates of all neurons in response to a given sample. Contrary to the wide range of firing rates observed per neuron, the firing rates of different samples are predominantly clustered around 0.25. This suggests that each input sample demands a relatively consistent and minimal number of firing spikes. The combination of high neuronal sparsity and consistent sample-level firing rates indicates that our Spiking-FullSubNet model is well-suited for deployment on energy-constrained neuromorphic hardware, as it can achieve significant computational and energy savings without compromising performance.

\section{Conclusion}
\label{sec:conclusion}
This paper introduces Spiking-FullSubNet, an SNN-based model architecture designed for real-time, ultra-low-power speech enhancement tasks. Spiking-FullSubNet leverages a full-band and sub-band fusion approach to effectively capture global and local spectral features crucial for speech enhancement. It incorporates two novel features: 1) a GSN spiking neuron model capable of capturing multi-scale temporal information, and 2) an efficient sub-band frequency partitioning approach that mimics human auditory perception. Our experimental results demonstrate superior speech enhancement performance with a nearly three orders of magnitude reduction in power consumption compared to SOTA ANN models. By leveraging the algorithmic advancements of Spiking-FullSubNet, this work presents a promising denoising solution for a wide range of audio devices. In future work, we will focus on implementing the proposed model on neuromorphic computing chips, such as Intel Loihi and Tianjic, to fully realize the potential of neuromorphic speech enhancement technologies.


\bibliography{main}
\bibliographystyle{IEEEtran}

\end{document}